\documentclass[12pt]{article}
\usepackage{amsmath,amsfonts,epsfig,mathrsfs,todonotes,yfonts}
\usepackage{calligra}
\usepackage{dsfont}
\usepackage{stmaryrd}

\usepackage{hyperref}
 \hypersetup{
     colorlinks=true,
     linkcolor=black,
     filecolor=back,
     citecolor=blue,      
     urlcolor=cyan,
     }
\usepackage[most]{tcolorbox}

\numberwithin{equation}{section}

\newcommand{\al}{\alpha}

\newcommand{\de}{\delta}

\newcommand{\OmegaB}{{l\!\!\! {\Omega}}}

\newcommand{\one}{{\mathds{1}}}

\newcommand{\nn}{\nonumber}
\newcommand{\na}{\nabla}

\newcommand{\ber}{\begin{eqnarray}}
\newcommand{\eer}[1]{\label{#1}\end{eqnarray}}
\newcommand{\eero}{\end{eqnarray}}

\newcommand{\half}{{\textstyle{\frac12}}}

\newcommand{\bbX}[1]{\mathbb{X}^{#1}}
\newcommand{\bbY}[1]{\mathbb{Y}^{#1}}

\newcommand{\bbD}[1]{\mathbb{D}_{#1}}
\newcommand{\bbDB}[1]{\bar{\mathbb{D}}_{#1}}
\def\+{{+\!\!\!+}} 
\def\pp{\mbox{\tiny${}_{\stackrel\+ =}$}}
\newcommand{\pa}[1]{\partial_{#1}}

\newcommand{\re}[1] {(\ref{#1})}

\DeclareMathAlphabet{\mathcalligra}{T1}{calligra}{m}{n}
\DeclareFontShape{T1}{calligra}{m}{n}{<->s*[2.2]callig15}{}
\newcommand{\scriptr}{\mathcalligra{r}\,}

\newcommand{\auth}
{\ \large Ulf Lindstr\"om ${}^{a,b}$\footnote{email: ulf.lindstrom@physics.uu.se} }
\begin{document}  
\hfill UUITP-33/22 
%\hfill\today 
\vspace{1cm}
                    \begin{center}
{\Large{\bf Yano $F$ structures and extended Supersymmetry.}}\\
\vspace{0.75cm}
\auth
\end{center}
\vspace{.5cm}
\centerline{${}^a${\it \small Department of Physics, Faculty of Arts and Sciences,}}
\centerline{{\it \small Middle East Technical University, 06800, Ankara, Turkey}}
\vspace{.5cm}
\centerline{${}^b${\it \small Department of Physics and Astronomy, Theoretical Physics, Uppsala University}}
\centerline{{\it \small SE-751 20 Uppsala, Sweden }}
\bigskip

{\bf Abstract}: \vspace{0,5cm}

It is shown how extended supersymmetry realised directly on the $(2,2)$ semichiral superfields of a symplectic sigma model gives rise to a geometry on the doubled tangent bundle consisting of two Yano $F$ structures on {an almost}  para-hermitian manifold. Closure of the algebra and invariance of the action is discussed in this framework and integrability of the $F$ structures is defined and shown to hold. The  reduction to the usual $(1,1)$ sigma model description and identification with 
the bi-quaternionic set of complex structures and their properties is elucidated. The $F$ structure formulation should be applicable to many other models and  will have an equivalent formulation in Generalised Geometry.
\eject

\tableofcontents
\section{Introduction}
Supersymmetric sigma models with varying numbers of  extra (non manifest) supersymmetries are legion. The number of supersymmetries determines the geometry of the target space as known ever since the papers by Zumino \cite{Zumino:1979et}, Alvarez-Gaume and Freedman, \cite{Alvarez-Gaume:1980xat}, Gates, Hull and Ro\v cek, \cite{Gates:1984nk} and by Hull and Witten \cite{Hull:1985jv}. From the $(1,1)$ superspace point of view finding the additional supersymmetries entails an ansatz containing tensors that closure of the algebra force to be complex structures. The extra supersymmetries may concern only a subset of the superfields as in \cite{Hull:2017hfa}, but the derivation is basically the same. Starting instead from a $(2,2)$ superspace formulation, the procedure typically first involves a reduction to $(1,1)$ to bring out the sigma model feature, a procedure that in some cases also involves eliminating auxiliary fields \cite{Buscher:1987uw}, \cite{Lindstrom:1994mw}. In those cases we have the option of asking about additional supersymmetries {\em before} the reduction and reveal a different geometric structure. This is what the present contribution is all about.

The model under investigation is the symplectic sigma model, which is a sigma model based only on semichiral superfields \cite{Buscher:1987uw}. Eliminating the auxiliary $(1,1)$ superfields is necessary for finding the sigma model geometry in terms of metric, complex structures, torsionfull connection etc on the tangent bundle. Realising extended supersymmetry directly as transformations of the $(2,2)$ semichiral superfields instead leads to  Yano $F$ structures on the doubled tangent bundle. From the $(1,1)$ point of view, this does not introduce new geometric objects, but identifies combinations of complex structures acting on subspaces defined by projection operators.

The presentation first gives a general background in Sec.~\!2, then, in Sec.~\!3, lists some results from the literature that are used in the main discussion in Sec.~\!4. Before that discussion, a simpler example is introduced in Sec.~\!3, to illustrate the situation.
 Sec.~\!5 contains a discussion of integrability of the $F$ structures on the  doubled tangent bundle and Sec.~\!6 relates the new geometry to the known bi-quaternionic structures and geometry in $(1,1)$ superspace. Sec.~\!7, finally contains some thoughts on the results and the general problem. Index conventions are summarised in Appendix B.

\section{Background}
Additional symmetries of a supersymmetric non linear sigma model are associated with additional complex structures. For $(2,2)$ models in $D=2$, the complex structures typically become manifest in the reduction to $(1,1)$ superspace, but are sometimes there already at the $(2,2)$ level. This is so for a $(2,2)$ sigma model  written in terms of chiral superfields $\phi : \bbDB{\pm}\phi=0$
\ber
S=\int d^2xd^2\theta d^2\bar\theta K(\phi , \bar \phi)~.
\eer{Achir1}
When $K$ is hyperk\"ahler, it has additional supersymmetry \cite{Hull:1985pq} with 
\ber
\de\phi^i=\bbDB{}^2\big(\bar\epsilon J^i(\phi,\bar\phi)\big)
\eer{}
provided that $\bbDB{\al}\epsilon=\bbD{}^2\epsilon=\pa{a}\epsilon=0$~. The complex structure is
\ber\nn
\mathbb{J}^{(1)} :=\left(\begin{array}{cc}0 &J^i,_{\bar j}\\ \bar J^{\bar i},_{ j}&0\end{array}\right)~.
\eer{Cst1}
The model now has $(4,4)$ supersymmetry with $\mathbb{J}^{(3)}$ the canonical complex structure present in $(2,2)$ and  
$\mathbb{J}^{(2)} =\mathbb{J}^{(3)}\mathbb{J}^{(1)}$ the remaining complex structure. The algebra closes modulo the field equations.\\

If we include twisted chiral superfields, $\chi: \bbDB{+}\chi= \bbD{-}\chi=0$, the model 
\ber
S=\int d^2xd^2\theta d^2\bar\theta K(\phi, \bar \phi, \chi,\bar \chi)~.
\eer{Achirtchir1}
can have a linear off-shell supersymmetry in terms of the $(2,2)$ fields, provided that
\ber\nn
&&K,_{\phi^i \bar\phi^j}+K,_{\chi^j \bar\chi^i}=0\\[1mm]
&&K,_{\phi^i \bar\phi^j}-K,_{\phi^j \bar\phi^i}=0~.
\eer{}
and that the complex structures that are covariantly constant wrt  connections with torsion,
as shown in  \cite{Gates:1984nk}.\\

The situation is a bit more complicated for models that depend on left and right semichiral superfields \cite{Buscher:1987uw}. The left $L=(\ell^a,\bar\ell^{\bar a})^t$ and right $R=(\scriptr^{a'},\bar{\scriptr}^{\bar a'})^t$ semichiral fields obey $\bbDB{+}\ell^a=0$ and  $\bbDB{-}\scriptr^{a'}=0$ repectively. To determine a nonlinear sigma model
\ber
S=\int d^2xd^2\theta d^2\bar\theta K(L,R)~,
\eer{Asemis1}
it must contain an equal number $d$ of left and right semichirals. The target space is then $4d$ real dimensional. It is shown in 
\cite{Goteman:2009xb} and further elaborated on in \cite{Lindstrom:2014bra} that $d=1$, i.e., four real dimensions, does not allow additional off-shell supersymmetry realised as transformations on the semichirals and in \cite{Goteman:2012qk} that the on-shell case implies hyperk\"ahler geometry on the target space\footnote{Exactly how these two results are reconciled for the $SU(2)\times U(1)$ model \cite{lost}, is discussed in \cite{Lindstrom:2014bra}}. The general case $d>1$ is treated both on and off-shell in \cite{Goteman:2009ye}. There off-shell supersymmetry is shown to lead to a surprising geometric structure. It is that discussion that is the starting point for the present paper.

\section{The crux of the matter}

The model \re{Asemis1} does not represent a sigma model in terms of the lowest field components. The target space geometry becomes apparent only after it is reduced to $(1,1)$ superspace and spinorial auxiliary fields are integrated out. We reduce using
\ber
\bbD{\pm}=D_\pm-iQ_\pm
\eer{Ddef}
we have 
\ber\nn
&&Q_+L=JD_+L~,~~~Q_-L=:\Psi_-\\[1mm]
&&Q_-R=JD_-R~,~~~Q_+R=:\Psi_+~,
\eer{Cdef}
where $J=diag(i,-i)$ is the canonical complex structure. Indices are suppressed in \re{Cdef}. When using indices they are assigned according to the paragraph above \re{Asemis1}. The non dynamical auxiliary fields $\Psi_\pm$ can be integrated out to yield the complex structures according to (for the plus transformations)
\ber\nn
\delta\left(\begin{array}{l}
L\\
R\end{array}\right) &=&\epsilon^+Q_+\left(\begin{array}{l}
L\\
R\end{array}\right)=\epsilon^+\left(\begin{array}{l}
JD_+L\\
\Psi^R_+\end{array}\right)\\[1mm]
&\to &J_{(+)}\epsilon^+D_+\left(\begin{array}{l}
L\\
R\end{array}\right)
\eer{Elim}
where  
\ber
\Psi^R_+=J^R_{(+)i}D_+X^i=J^R_{(+)i}D_+\bbX{i}_|~,
\eer{Psieq+}
$J_{(+)}=J_{(+)}(L,R)$, and $i$ runs over all indices. Similarly we find 
\ber\Psi^L_-=J^L_{(-)i}D_-\bbX{i}_|~.
\eer{Psieq-}
The discussion of extended supersymmetries now proceeds from the $(1,1)$ sigma model formulation and results in a bi-$SU(2)$ structure for $(4,4)$ symmetry, as described in \cite{Gates:1984nk}. 

If we ask for additional symmetries realised on $(L,R)$ {\em before} the reduction we will clearly not find $J_{(\pm)}$. The following example illustrates the situation.

\subsection{Susy on right semichirals only}
\label{right}

As an example and illustration we first sketch the situation when supersymmetry realised in the right sector only.
A consistent transformation of the right semichirals $R$ leaving the $L$ sector invariant is
\ber
\bar\delta \left(\begin{array}{l}
L\\
R\end{array}\right) =
\bar \epsilon ^+\left( \begin{array}{cccc}
0 &0&0&0\cr
0&0&0&0\cr
0 &0&    g^{a'}_{b'}&0\cr
0 &0&0&-\bar{    h}^{\bar a'}_{\bar b'}\end{array}\right)\bbDB{+}\left(\begin{array}{l}
L\\
R\end{array}\right) =:U \bar\epsilon ^+\bbDB{+}\left(\begin{array}{l}
L\\
R\end{array}\right) 
\eer{Rtrans}
Introducing $\bbX{} :=(L^a,R^{a'})^t$, this gives 
\ber
\delta \bbX{}=U \bar\epsilon ^+\bbDB{+}\bbX{}+V \epsilon ^+\bbD{+}\bbX{}~,
\eer{gtrans}
where
\ber
V =\left( \begin{array}{cccc}
0 &0&0&0\cr
0&0&0&0\cr
0 &0&-    h^{a'}_{b'}&0\cr
0 &0&0&\bar{    g}^{\bar a'}_{\bar b'}\end{array}\right)
\eer{}
is the complex conjugate of $U $ rearranged to act on $\bbX{}$. Closure of the algebra follows if
\ber\nonumber
U V =-\textrm{diag}(0,0,\one,\one)=V U 
\eer{info}
i.e., if $    g^{a'}_{b'}    h^{b'}_{c'}=\delta^{a'}_{c'}$. Clearly $U $ and $V $ are degenerate and do not separately square to minus one. They do not combine into a complex structures, but they have a geometric interpretation as a Yano $F$-structure on a doubled space. For a manifold ${\cal M}$ such a structure  is a map $f$ of its tangent bundle $T{\cal M}$ to itself \cite{Yano:1961} :
\ber
f: T{\cal M} \to T{\cal M}~,~~~f^3+f=0~.
\eer{fstr}
This condition is a generalisation of the condition $J^2=-1$ for an almost complex structure and allows for degenerate matrices.
In the present case we see that  the $8d\times 8d$ matrix
\ber
{\cal F}:=\left( \begin{array}{cc}
0 &U \cr
V &0
\end{array}\right)
\eer{}
defined on the double tangent bundle  $T{\cal M}\oplus T{\cal M}$ satisfies \re{fstr}. The transformations \re{gtrans} now read
\ber
\delta {\boldsymbol X}:=\left(\begin {array}{c}
\bar\delta \bbX{}\cr
\delta \bbX{}\end{array}\right)={\cal F}\left(\begin {array}{c}
\epsilon^{+}\bbD{+}\bbX{}\cr
\bar\epsilon^{+}\bbDB{+}\bbX{}\end{array}\right)~,
\eer{dtrans}
and we may take the tangent bundles to have coordinates  $(\bbX{}, \bbD{+}\bbX{})$ and $(\bbX{}, \bbDB{+}\bbX{})$ respectively, so that the doubled tangent bundle has coordinates $(\bbX{}, \bbD{+}\bbX{},  \bbDB{+}\bbX{})$.

An $F$ structure defines projection operators. With all submatrices of dimension $2d\times 2d$, we
have that
\ber
l:=-{\cal F}^2=\left( \begin{array}{cccc}
0 &0&0&0\cr
0&\one&0&0\cr
0 &0&0&0\cr
0 &0&0&\one\end{array}\right) \makebox{and} 
~m:=\one+{\cal F}^2=\left( \begin{array}{cccc}
\one &0&0&0\cr
0&0&0&0\cr
0 &0&\one&0\cr
0 &0&0&0\end{array}\right)
\eer{}
obey
\ber\nn
&&	l   + m   = \one,\quad
	l^2   = l  , \quad m^2   = m  ~,\quad l   m =ml  = 0 \\[1mm]
&&
	{\cal F}   l   = l   {\cal F}   = {\cal F}  , \quad m  {\cal F}   ={\cal F}   m  = 0~.
\eer{fund1}
Clearly $l$ projects onto the $4d$ subspace corresponding to $R$ in the doubled tangent space and $m$ to its complement, i.e., the $L$ subspace. 
{Projection by $l$ makes $l{\cal F}l $  an almost complex structure on the $4d$ subspace corresponding to $R$:
\ber
(l{\cal F}l )^2=-l~.
\eer{} }
Once we show that $U$ and $V$ are integrable, it follows that ${\cal F}$ is a complex structure on the $R$ subbundle projected by $l$, and that this subbundle is involutive wrt to the corresponding projection operator. This implies that we have an alternative characterisation of  ${\cal F}$ and the subbundle as a CR structure. We return to these issues in more detail in the context of the general conditions for supersymmetry of semichirals below.

As a final comment on this example we discuss invariance of the action under the transformations \re{Rtrans}. We have
\ber
\delta S=\int d^2xd^2\theta d^2\bar\theta K_{a'}(L,R)\big(g^{a'}_{b'}\bar \epsilon^+\bbDB{+}\bbX{b'}- h^{a'}_{b'}\epsilon^+\bbD{+}\bbX{b'}\big)~.
\eer{}
Focussing on the $\bar\epsilon$ transformation and pushing in $\bbDB{+}$ from the measure we must have
\ber
\int d^2xd^2\theta \bbDB{-} \Big(K,_{a' i}\bbDB{+}\bbX{i}g^{a'}_{b'}+K_{a' }g^{a'}_{b'},_{c'}\bbDB{+}\bbX{c'}\Big)\bar \epsilon^+\bbDB{+}\bbX{b'}=0~,~~i\ne a~.
\eer{}
The vanishing of the terms in parenthesis is necessary and sufficient for $\bar\epsilon$ invariance.
\section{Extended susy realised on semichirals}
In this section we consider the possible realisation of two additional left and two additional right complex  supersymmetries  on an equal number of right and left semichirals.  The derivation will be a generalisation of the discussion of the example in subsection \ref{right}, and will fill in the details left out there.  
\subsection{Results from Ref. \cite{Goteman:2009ye}. }
\label{GLRR}

The starting point is thus again the action \re{Asemis1} in terms of the generalised potential $K(L,R)$. Replacing \re {gtrans}, the general ansatz for the transformations of $\bbX{i}=(L,R)^t$ compatible with the chirality constraints $\bbDB{+}\ell=0$ and  $\bbDB{-}\scriptr=0$ is:
\ber
&\bar\delta^{(\al)}\bbX{i}+\delta^{(\al)}\bbX{i}:=
\bar\epsilon^\alpha U^{(\alpha)i}{}_j\bbDB{\alpha}\bbX{j}+\epsilon^\alpha V^{(\alpha)i}{}_j\bbD{\alpha}\bbX{j}
\eer{UVdef}
where $\al = (\pm)$, so two new left and two new right transformations. The explicit form of the transformation matrices are
\ber
&U^{(+)}=\left( \begin{array}{cccc}
\ast &f^{a}_{\bar b}&f^{a}_{ b'}&f^{a}_{\bar b'}\cr
\ast &0&0&0\cr
\ast &0&g^{a'}_{b'}&0\cr
\ast &0&0&-\bar{h}^{\bar a'}_{\bar b'}\end{array}\right)
\quad
&U^{(-)}=\left( \begin{array}{cccc}
\tilde g^a_b		&		0		&\ast 		&0\cr
0				&-\bar {\tilde h}^{\bar a}_{\bar b}	&\ast 	&0\cr
\tilde f^{a'}_{ b}		&\tilde f^{a'}_{\bar b}	&\ast &\tilde f^{a'}_{\bar b'}\cr
0&0&\ast &0\end{array}\right)
\eer{MNdef2}
and
\ber
V^{(\pm)}=-\left(\begin{array}{cc} \sigma_1 & 0\cr
0 &\sigma_1\end{array}\right)\bar U^{(\pm)}\left(\begin{array}{cc} \sigma_1 & 0\cr
0 &\sigma_1\end{array}\right)
\eer{Ndef}
where lower indices on the functions $f$  denote derivatives and where $\sigma_1$ is the first Pauli matrix times the appropriate unit matrix. The stars denote the matrix elements that do not enter \re{UVdef}. In this paper they are set to zero.\\

In \cite{Goteman:2009ye} we derive the following from the  {\em closure of the algebra} :
\begin{enumerate}
\item The transformation matrices  $U^{(\pm)}$ commute as do $V^{(\pm)}$.
 \item The products $U^{(\pm)}V^{(\pm)}$ and $V^{(\pm)}U^{(\pm)}$ equal minus one except for a zero row\footnote{This is not possible for just one left and one right semichiral. Hence the exclusion of $d=1$. Pseudo supersymmetry is possible, however. }.
\item The transformation matrices are all separately integrable.
\item The Magri-Morosi concomitant vanishes for all two pairs of the transformation matrices.  
\end{enumerate}
Together with the previous, the fourth point leads to the  fact that the matrices $U^{(\al)}$ (and $V^{(\al)}$) are degenerate and satisfy
\ber\nonumber
&&U^{(+)}V^{(+)}=-\textrm{diag}(\one,0,\one,\one), \quad V^{(+)}U^{(+)}=-\textrm{diag}(0,\one,\one,\one),\\[1mm]
&&U^{(-)}V^{(-)}=-\textrm{diag}(\one,\one,\one,0), \quad V^{(-)}U^{(-)}=-\textrm{diag}(\one,\one,0,\one)
\eer{info}
again prevents a direct interpretation in terms of complex structures on the tangent space $T{\cal M}$.
And, as in the example in subsection \ref{right}, we are led to consider endomorphisms on the doubled tangent bundle\footnote{Not to be confused with the second tangent bundle.} (Whitney sum) $T_{(\al)}{\cal M}\oplus \bar T_{(\al)}{\cal M}$ and  $F$-structures instead. (Here bar is not complex conjugation, it only denotes one of the bundles.) The following $8d\times 8d$ matrices are $F$-structures\footnote{For pseudo supersymmetry  \re{fst} is replaced by $ \tilde{\cal F}^{3}_{(\al)}-\tilde{\cal F}_{(\al)}=0$.} in the sense of Yano
\cite{Yano:1961}:
\ber
{\cal F}_{(\al)}:=\left(\begin {array}{cc}
0&U^{(\al)}\cr
V^{(\al)}&0\end{array}\right)\quad \implies {\cal F}^{3}_{(\al)}+{\cal F}_{(\al)}=0~.
\eer{fst}
 In terms of $ {\cal F}_{(\al)}$ we may write the transformations \re{UVdef} as
\ber
\left(\begin {array}{c}
\bar\delta^{(\alpha)}\bbX{}\cr
\delta^{(\alpha)}\bbX{}\end{array}\right)
={\cal F}_{(\alpha)}\left(\begin {array}{c}
\epsilon^{(\alpha)}\bbD{\alpha}\bbX{}\cr
\bar\epsilon^{(\alpha)}\bbDB{\alpha}\bbX{}
\end{array}\right)~,
\eer{YFtfs}
in analogy to how complex structures act.
%Moreover,  $-{\cal F}^2_{(\pm)}$ and $1+{\cal F}^2_{(\pm)}$ define integrable distributions \rd{\tt On $T_{(\alpha)}M\oplus \bar T_{(\alpha)}M$??  Recheck!}. 
We define 
\ber
\hat P_{   \al}	= \one + V^{(   \al)} U^{(   \al)}  ~,\quad
P_{   \al}		= \one + U^{(   \al)}V^{(   \al)} ~,
\eer{projs}
and
\begin{equation}
	m_{(   \al)}:=\one+{\cal F}^{2}_{(   \al)} = \left(
	\begin{array}{cc}
		P_   \al & 0 \\ 0 & \hat P_   \al
	\end{array}\right), \quad
	l_{(   \al)}:=-{\cal F}^{2}_{(   \al)} = \left(
	\begin{array}{cc}
		\one-P_   \al & 0 \\ 0 & \one-\hat P_   \al
	\end{array}\right).\label{projops}
\end{equation}
The conditions on $U^{(\alpha)}$ and $V^{(\alpha)}$ may  then be used to show that these fulfil
\ber
	l_{(   \al)} + m_{(   \al)} = \one,\quad
	l^2_{(   \al)} = l_{(   \al)}, \quad m^2_{(   \al)} = m_{(   \al)}~,\quad l_{(   \al)} m_{(   \al)} =0
\eer{fund1}
and
\ber
	{\cal F}_{(   \al)} l_{(   \al)} = l_{(   \al)} {\cal F}_{(   \al)} = {\cal F}_{(   \al)}, \quad m_{(   \al)}{\cal F}_{(   \al)} ={\cal F}_{(   \al)} m_{(   \al)} =0.
\eer{fund2}
The operators $l_{(   \al)}$ and $m_{(   \al)}$ applied to the tangent space at each point of the manifold are complementary projection operators and define complementary distributions $\Lambda_   \al$, the first fundamental distribution, and $\Sigma_   \al$, the second fundametal distribution, corresponding to $l_   \al$ and $m_   \al$, of dimensions $6d$ and $2d$, respectively. 

Let $\mathcal{N}_{{\cal F}_{(\al)}}$ denote the Nijenhuis tensor for the $F$-structures ${\cal F}_{(\al)}$. By a theorem of Ishihara and Yano \cite{IshiharaYano} we have that
\begin{enumerate}
	\item[i.]
	$\Lambda_\al$ is integrable iff $m^i_{(\al)l} \mathcal{N}_{\mathcal{F}_{(\al)} jk}^l=0$,
	\item[ii.]
	$\Sigma_\al$ is integrable iff $\mathcal{N}_{\mathcal{F}_{(\al)} jk}^i m^j_{(\al)l} m^k_{(\al) m} =0$.
\end{enumerate}
From the definition of the $F$-structures in terms of $U$ and $V$ in \re{fst} {and points 3 and 4 in the list above}, one may argue that these two conditions are fulfilled. This indeed follows from the discussion of integrability in Sec. \ref{Fint} below. Hence, the distributions $\Lambda_\al$ and $\Sigma_\al$ are integrable. \\

\subsection{Invariance of the action}
In this section we develop the conditions for invariance of the action briefly touched upon in \cite{Goteman:2009ye}.

In the special case when all entries in the $U$ and $V$ matrices are derivatives of function,\footnote{This is the case for three of the five non-zero entries, but  two need not be gradients.} the condition from 
 {\em invariance of the action in \re{Asemis1}} for the $\bar\epsilon^{\al} $ transformations in \re{UVdef} now reads:
\begin{equation}
\label{hermiticity}
0=\left(K,_{i}U^{(\al)i}{}_{[j}\right){}_{k]}=K,_{i[j}U^{(\al)i}{}_{k]}~, 
\end{equation}
with $i\ne \bar a, ~~ (j,k) \ne (b,c)$ for $\al=+$ and $i\ne \bar a', ~~ (j,k) \ne (b',c')$ for $\al=-$. The conditions for invariance under $\epsilon^{\al} $ transformations follow by complex conjugation, replacing $U\to V$.
The consequence of this may be formulated in terms of the antisymmetric tensor ${\OmegaB}$ which maps $ TM\oplus \bar TM$ to $\bar T^*M\oplus  T^*M$ and is defined as 
\ber
 {\OmegaB} =\left(\begin{array}{cc}
0 & \mathbb{K}\\ - \mathbb{K}^t & 0
\end{array}\right)~,
 \eer{omg}
 with $\mathbb{K}$ being the Hessian of the potential $K(L,R)$.

Consider the projection of ~${\OmegaB}$ to the subspace defined by $l_{(+)}$. From \re{projs}, \re{projops} we need
\ber
(UV)^{(+)}\mathbb{K}(VU)^{(+)}&&=
\left(\begin{array}{cccc}
0 & K,_{a\bar b}&K,_{ab'}&K,_{a\bar b'}\\  
0& 0&0&0\\
0&K,_{a'\bar b}&K,_{a'b'}&K,_{a'\bar b'}\\
0&K,_{\bar a'\bar b}&K,_{\bar a'b'}&K,_{\bar a'\bar b'}
\end{array}\right)=:{\mathbb{K}}^{(+)}~.
\eer{tildeK+}
and
\ber
-(VU)^{(+)}\mathbb{K}^t(UV)^{(+)}=-\Big((UV)^{(+)}\mathbb{K}(VU)^{(+)}\Big)^t=\Big({\mathbb{K}}^{(+)}\Big)^t~.
\eer{}
Examining \re{hermiticity} for $\al = +$ shows that ${\mathbb{K}}^{(+)}$ is precisely the  matrix that enters this case.
Similarly we find 
\ber
(UV)^{(-)}\mathbb{K}(VU)^{(-)}&&=
\left(\begin{array}{cccc}
K,_{ab} & K,_{a\bar b}&0&K,_{a\bar b'}\\  
K,_{\bar ab}& K,_{\bar a\bar b}&0&K,_{\bar a\bar b'}\\
K,_{a' b}&K,_{a'\bar b}&0&K,_{a'\bar b'}\\
0&0&0&0
\end{array}\right)=:{\mathbb{K}}^{(-)}~,
\eer{tildeK-}
the matrix that enters the $\al = -$ relation in \re{hermiticity}. Combining the results from  \re{tildeK+} to \re{tildeK-} we write the conditions for invariance as
\ber
[{\mathbb{K}}^{(\al)},U^{(\al)}]=0~,~~\makebox{and c.c.}~,
\eer{inv}
and the projections of ${\OmegaB}$ as
\ber
l_{(\al)}~\!{\OmegaB}~\!l_{(\al)}=:{\OmegaB}^{(\al)}~,
\eer{}
where  ${\mathbb{K}}\to {\mathbb{K}}^{(\al)}$ on the right hand side. Using \re{fund2} we then find
\ber
\mathcal{F}^t_{(\al)}{\OmegaB}^{(\al)}\mathcal{F}_{(\al)}={\OmegaB}^{(\al)}~,
\eer{}
i.e. that the $F$-structures preserve ${\OmegaB}$ on the subspaces $\Lambda^{(\al)}$ defined by the projectors $l_{(\al)}$ .

As a two-form ${\OmegaB}$  satisfies $d~\!{\OmegaB}=0$;
\ber\nn
d~\!{\OmegaB}&=&2d\Big(K,_{ij}d\bar z^i \wedge dz^j \Big)\\[1mm]&=&2\Big(K,_{ijk}d\bar z^k\wedge d\bar z^i \wedge dz^j +K,_{ijk}dz^k\wedge d\bar z^i \wedge dz^j \Big)=0~,
\eer{domg} 
where bar and unbarred $z$ denote coordinates on the  two copies of the tangent bundle, and the last equality follows by a symmetric contracted into an antisymmetric tensor. So ${\OmegaB}$ it is a symplectic form on the doubled tangent space, as long as it is nowhere vanishing. 

\section{Doubled tangent bundle and {almost} para-hermitian structure.}
We have taken the doubled tangent bundle to be defined according to the Whitney sum of two tangent bundles \cite{Whitney} and used it informally. In particular we have not addressed the issue of integrability for endomorphisms on $T_{(\al)}{\cal M}\oplus \bar T_{(\al)}{\cal M}$. Below we introduce a doubled manifold $\mathbb{M}={\cal M}^2$ and  integrability of endomorphisms of $T\mathbb{M}$ via the Lie bracket. The doubled tangent space then follows from a double foliation of $\mathbb{M}$, the identification  of coordinates on  $\mathbb{M}$ as coordinates on two  leafs  $S$ and $S^\star $of the foliation and finally the restriction to the physical ${\cal M}$ as a base via a section condition.
\subsection{Doubling the base space}
This section parallels the discussion of Double Field theory, as presented, e.g., in \cite{Kimura:2022jyp}.

So we  start from doubling the manifold ${\cal M} \to \mathbb{M}$ where $\mathbb{M}$ has twice the dimension of ${\cal M}$, i.e., ( $4d \to 8d$ real dimensions). 
The symplectic form $\OmegaB$ in \re{omg} can be thought of as a the product of two matrices on this doubled space
\ber
 {\OmegaB} =\left(\begin{array}{cc}
 \mathbb{K} &0\\  0& -\mathbb{K}^t
\end{array}\right)
\left(\begin{array}{cc}0&\one_{4d}\\\one_{4d}&0\end{array}\right):={{\cal K}\eta}~,
\eer{}
where 
\ber
{\eta}^2=\left(\begin{array}{cc}0&\one_{4d}\\\one_{4d}&0\end{array}\right)^2=\one_{8d}~,
\eer{}
and ${\cal K}$ is a neutral metric of signature $(4d,4d)$.
These conditions identify $(\mathbb{M},{\OmegaB},{{\eta}})$ as an almost para-hermitian manifold.

Now $\eta$ is a local product structure corresponding to the doubling of the tangent bundle. It  introduces two distributions $L$ and $ L^\star$ via the projection operators
\ber
P=\half\Big(\one+{\eta}\Big)~,\qquad \tilde P=\half\Big(\one-{\eta}\Big)~.
\eer{}
Using these we have
\ber
T\mathbb{M}=L\oplus L^\star~,
\eer{}
where $L=PL$ and $L^\star=\tilde PL^\star$ are the $+1$ and $-1$  eigen spaces of ${\eta}$.

Given one of these, $L$ say, it is almost immediate that it is involutive since the projection operators are constant. We have that  $V,W\in L$ implies
\ber\nn
&&\llbracket \mathbb{V},\mathbb{W}\rrbracket^j_{\cal L}=P^i_{k}P^j_{m}\big(V^k\partial_iW^m-W^k\partial_iV^m\big)\\[1mm]
\Rightarrow &&\tilde P^s_j\llbracket \mathbb{V},\mathbb{W}\rrbracket^j_{\cal L}=
 \tilde P^s_jP^j_{m}P^i_{k}\big(V^k\partial_iW^m-W^k\partial_iV^m\big)=0~.
\eer{}
By the Frobenius theorem, the involutive bundle $L$ defines a foliation structure in $\mathbb{M}$ so that
$L = TS$. The  spacetime ${\cal M}$  is identified with a leaf of the foliation. With
 the basis of $L $ as $\partial_i=\frac \partial {\partial \bbX{i}}$,  the local coordinates of  $S$  are $\bbX{i}$. Similarly,  for $L^\star$  the local coordinates on $S^\star$ are $\bbY{i}$. The pair $(S,S^\star)$ defines a doubled foliation in $\mathbb{M}$ and correspond to the coordinate system\footnote{These coordinates may be identified with the coordinates $(z,\bar z)$ used in \re{domg}.}
$(\bbX{i},\bbY{j})$. 

\subsection{Integrability of  $({\cal F})$}
\label{Fint}
As discussed in, \cite{IshiharaYano}, the vanishing of the Nijenhuis tensor
\ber
{\cal N}_{IJ}^{~~P}({\cal F})=-{\cal F}^P_K\partial_{[I}{\cal F}^K_{J]}+{\cal F}^K_{[I}\partial_{|K|}{\cal F}^P_{J]}=0~
\eer{Nij}
ensures integrability. We thus consider, 
\ber\nn
&&{\cal N}_{IJ}^{~~X}=-{\cal F}^X_Y\partial_{[I}{\cal F}^Y_{J]}+{\cal F}^K_{[I}\partial_{|K|}{\cal F}^X_{J]}\\[1mm]
&&{\cal N}_{IJ}^{~~Y}=-{\cal F}^Y_X\partial_{[I}{\cal F}^X_{J]}+{\cal F}^K_{[I}\partial_{|K|}{\cal F}^Y_{J]}~,
\eer{Nij2}
where $X$ and $Y$ denote arbitrary $\bbX{i}$ and $\bbY{j}$ indices and $I,J,P,...$ runs over both. 
It follows that 
\ber\nn
&&{\cal N}_{XX}^{~~X"}=-{\cal F}^{X"}_Y\partial_{[X}{\cal F}^Y_{X]}+{\cal F}^K_{[X}\partial_{|K|}{\cal F}^{X"}_{X]}
=-{\cal F}^{X"}_Y\partial_{[X}{\cal F}^Y_{X]}\\[1mm]\nn
&&{\cal N}_{XY}^{~~X"}=-{\cal F}^{X"}_Y\partial_{[X'}{\cal F}^Y_{Y']}+{\cal F}^K_{[X'}\partial_{|K|}{\cal F}^{X"}_{Y']}=\half \Big({\cal F}^{X"}_Y \partial_{Y'}{\cal F}^Y_{X'}+{\cal F}^Y_{X'}\partial_{Y}{\cal F}^{X"}_{Y'}\Big)\\[1mm]\nn
&&{\cal N}_{YY}^{~~X"}=-{\cal F}^{X"}_Y\partial_{[Y}{\cal F}^Y_{Y]}+{\cal F}^K_{[Y}\partial_{|K|}{\cal F}^{X"}_{Y]}
={\cal F}^{X'}_{[Y}\partial_{|X'|}{\cal F}^{X"}_{Y]}
\\[1mm]\nn
&&{{\cal N}_{XX}^{~~Y}}=-{\cal F}^Y_{X'}\partial_{[X}{\cal F}^{X'}_{X]}+{\cal F}^K_{[X}\partial_{|K|}{\cal F}^Y_{X]}
={\cal F}^{Y'}_{[X}\partial_{|Y'|}{\cal F}^Y_{X]}\\[1mm]\nn
&&{\cal N}_{XY}^{~~Y"}=-{\cal F}^{Y"}_{X'}\partial_{[X}{\cal F}^{X'}_{Y]}+{\cal F}^K_{[X}\partial_{|K|}{\cal F}^{Y"}_{Y]}
=\half \Big({\cal F}^{Y"}_{X'} \partial_{X}{\cal F}^{X'}_{Y}+{\cal F}^{X'}_{Y}\partial_{X'}{\cal F}^{Y"}_{X}\Big)\\[1mm]
&&{\cal N}_{YY}^{~~{Y"}}=-{\cal F}^{Y"}_X\partial_{[Y}{\cal F}^X_{Y]}+{\cal F}^X_{[Y}\partial_{|X|}{\cal F}^{Y"}_{Y]}-=-{\cal F}^{Y"}_X\partial_{[Y}{\cal F}^X_{Y]}~.
\eer{}
To return to the doubled tangent bundle (over one copy of the base manifold ${\cal M}$) we must impose a section condition that identifies a section of $\mathbb{M}$ as the base space. We thus require $\partial_Y$ to vanish on all functions. So while the integrability condition for ${\cal F}$ on $T\mathbb{M}$ is ${\cal N}_{IJ}^{~~K}=0$, on the doubled tangent bundle this gets restricted to the vanishing of
\ber\nn
{\cal N}_{XX}^{~~X"}&=&\partial_{[X}{\cal F}^{X"}_{|Y|}{\cal F}^Y_{X]}-{\cal F}^{X"}_Y\partial_{[X}{\cal F}^Y_{X]} ~,\\[1mm]\nn
{\cal N}_{YY}^{~~X"}&=&{\cal F}^{X'}_{[Y}\partial_{|X'|}{\cal F}^{X"}_{Y]}\\[1mm]
{\cal N}_{XY}^{~~Y"}&=&\half \Big({\cal F}^{Y"}_{X'} \partial_{X}{\cal F}^{X'}_{Y}+{\cal F}^{X'}_{Y}\partial_{X'}{\cal F}^{Y"}_{X}\Big)~.
\eer{Nijlist}
After employing the section conditions, all indices can be replaced by the $(i,j,\dots)$ of the original formulation on $T{\cal M}$.
The ${\cal N}_{XX}^{~~X}$ conditions become
\ber
{\cal N}_{ij}^{~~m}&=&\partial_{[i}{U}^{m}_{|k|}{V}^{k}_{j]}=0~,
\eer{}
where the last equality is a consequence of  $VU=-\one$ and ${V}^{k}_{j}={V}^{k},_{j}$.
Similarly, the $ {\cal N}_{XY}^{~~Y}$ condition gives
\ber
{\cal N}_{ij}^{~~m}&=&\half \Big(V^{m}_{k} \partial_{i}U^{k}_{j}+U^{k}_{j}\partial_{k}V^{m}_{i}\Big)
=\half \partial_{i} \Big(V^{m}_{k}U^{k}_{j}\Big)=0~.
\eer{}
The condition from ${\cal N}_{YY}^{~~X"}$ gives
\ber
{\cal N}_{ij}^{~~m}&=&U^k_{[i}\partial_{|k|}U^{m}_{j]}=\tilde{\cal N}_{ij}^{~~m}(U)=0~,
\eer{}
where the last identity follows is needed  for closure of the algebra in the $T{\cal M}$ formulation.
We thus see that  the three conditions on ${\cal N}_{ij}^{~~m}$ that follow from\re{Nijlist} are satisfied due to the conditions for closure of the algebra in the original $T{\cal M}$ formulation,  Conversly these could be derived from the integrability of ${\cal F}$.

Note that the vanishing of the Nijenhuis tensor for $V$, also needed for closure, is not independent of other conditions, In fact we have
\ber\nn
\tilde {\cal N}_{ij}^{~~m}(V)= 0 &\iff&  -(\partial_kV^m_s)U_m^q-V^m_s(\partial_mU^q_k)=0\\[1mm]
&\iff&-\partial_k(V^m_sU_m^q)=0
\eer{}
and the last equivalence again holds due to $VU=-\one$~. 

The vanishing of the Nijenhuis tensor together with $d~\!\OmegaB=0$ makes the geometry para-K\"ahler.

\section{Relation to the  bi-quaternionic structures in $(1,1)$}
In this section we relate the $(2,2)$ transformations \re{} to the known $(1,1)$ picture with two quaternion structures mentioned  around \re{SU2} above.

It was shown in \cite{Gates:1984nk} from a $(1,1)$ superspace perspective that $(4,4)$ symmetry of a sigma model with torsion implies the existence of two quaternionic structures:
\ber
J^{({\cal A})}_{(\pm)}J^{({\cal B})}_{(\pm)}=-\delta^{{\cal A}{\cal B}}+\epsilon^{{\cal A}{\cal B}{\cal C}}J^{({\cal C})}_{(\pm)}~,
\eer{SU2}
covariantly constant wrt a connection with torsion formed from the three-form $H=dB$
\ber
\na^{(\pm)}_{\pm}J_{(\pm)}=(\na_{\pm}^0\pm T)J^{(\pm)}=(\na_{\pm}^0\pm \half Hg^{-1})J_{(\pm)}=0~,~~~
\eer{covj}
 and a metric $g$ is hermitean wrt all complex structures\footnote{Sometimes referred to as  bi-hypercomplex geometry. }. 
The metric and $B$-field read \cite{Lindstrom:2005zr}
\ber
g=\Omega [J_{(+)}, J_{(-)}]~,~~~B=\Omega \{J_{(+)}, J_{(-)}\}~.
\eer{MetB}
where
\ber\nn
\Omega :=\left(\begin{array}{cc}0 &2iK,_{LR}\\ -2iK,_{RL}&0\end{array}\right)
~,~~~d\Omega=0~.
\eer{Omega}
From the point of view of generalised geometry, i.e., in terms of geometry defined using the generalised tangent bundle $\mathbb{T}{\cal M}={T}{\cal M}\oplus{T^*}{\cal M}$, the geometry is generalised K\"ahler of symplectic type.

\subsection{Reduction}

Descending to $(1.1)$ superspace and eliminating the auxiliary spinors\footnote{Eliminating the $\Psi$s is only going partially on-shell, sometimes called $(2,2)$ shell, since the $X$ field equations are not satisfied}, $\Psi_-$ and  $\Psi_+$, we find from  \re{Ddef} and \re{Cdef} and \re{Psieq+} and \re{Psieq-}, that
\ber\nonumber
&&\bbD{\pm}\bbX{i}_|=\bar \pi^{i}_{(\pm)k}D_\pm \bbX{k}_|\\[1mm]
&&\bbDB{\pm}\bbX{i}_|=\pi^{i}_{(\pm) k}D_\pm \bbX{k}_|
\eer{proj}
where we have introduced the projection operators
\ber
\pi_{(\pm)}:=\half \left( \one + iJ_{(\pm)}\right)~,
\eer{projo}
and their complex conjugates.
As mentioned, from  the $N=(1,1)$  analysis of \cite{Gates:1984nk} we know 
that when the model has $(4,4)$ supersymmetry there exists two $SU(2)$ worth\footnote{For positive definite metric.} of left and right complex structures $J_{(\pm)}^{({\cal A})}=(J_{(\pm)}^{(1)},J_{(\pm)}^{(2)},J_{(\pm)}^{(3)})$ 
on the $4d$ dimensional space \re{SU2}.
The metric \re{MetB} is hermitean with respect to all these complex structures and these in turn are $(\pm)$-covariantly constant according to their labels.
This determines the bi-hypercomplex target space geometry.
We now relate the $F$-structures to $J_{(\pm)}^{({\cal A})}$.\\

\subsection{The nonmanifest supersymmetries in $(1,1)$}
Using \re{proj}, the infinitesimal transformations \re{UVdef} can now be written  (dropping the $\theta$-projection)
\ber
&\delta\bbX{i}:=
\bar\epsilon^\alpha U^{(\alpha)i}{}_j\pi^{j}_{(\alpha)k}D_{\alpha}\bbX{k}+\epsilon^\alpha V^{(\alpha)i}{}_j\bar \pi^{j}_{(\alpha)k}D_{\alpha}\bbX{k}
\eer{UVdefOne}
with $U$ and $V$ satisfying the conditions in subsection \ref{GLRR}.
A number of conditions follow from the existence of the $J_{(\pm)}^{({\cal A})}$s. The requirement that the $U^{(+)}$ and $V^{(+)}$ transformations commute with the existing transformations generated by $J_{(+)}=J_{(+)}^{(3)}$, and idem for $U^{(-)}$ and $V^{(-)}$, implies
\ber\nn
&&U^{(\alpha)i}{}_j\pi^{j}_{(\alpha)k}
=\bar\pi^{i}_{(\alpha)k}U^{(\alpha)k}{}_j\pi^j_{(\alpha)k}\\[1mm]
&&V^{(\alpha)i}{}_j\bar\pi^{j}_{(\alpha)k}
=\pi^{i}_{(\alpha)k}V^{(\alpha)k}{}_j\bar\pi^j_{(\alpha)k}
\eer{pUp}
This is also verified using the explicit forms \re{MNdef2} of $U^{(\pm)}$ and $V^{(\pm)}$ and the known explicit forms of $J_{(\pm)}$ found by combining \re{Elim} with \re{Psieq+} and \re{Psieq-}:
\ber
J_{(+)}= \left(\begin{array}{ll}
J&0\\
J_{(+)L}^R&J^R_{(+)R}
\end{array}\right)~~,~~~~
J_{(-)}= \left(\begin{array}{ll}
J_{(-)L}^L&J_{(-)R}^L\\
0&J
\end{array}\right)
\eer{Jpm}

Multiplication by $J_{(\pm)}^{(1)}$ and $J_{(\pm)}^{(2)}$ yields the relations
\ber\nn
&&\half \left(J_{(\alpha)}^{(1)}-iJ_{(\alpha)}^{(2)}\right)U^{(\alpha)}\pi_{(\alpha)} =0%U^{(\alpha)}\pi_{(\alpha)}
\\[2mm]
&&\half\left(J^{(1)}_{(\alpha)}+iJ^{(2)}_{(\alpha)}\right)V^{(\alpha)}\bar\pi_{(\alpha)}=0%V^{(\alpha)}\bar\pi_{(\alpha)}
\eer{Jmult}
From the results in  subsec. \ref{GLRR} it also follows that
\ber
VU\pi=-\pi~,~~~UV\bar\pi=-\bar\pi~,
\eer{VU}

The structure becomes clearer if we observe that in terms of the $J_{(\pm)}^{({\cal A})}$s the additional supersymmetries take the form
\ber
\delta_s\bbX{}:= \delta^{\pm}\bbX{}+\bar\delta^{\pm}\bbX{}=\half\left(\left(J^{(1)}_{(\pm)}+iJ^{(2)}_{(\pm)}\right)\epsilon^\pm D_\pm\bbX{}+\left(J^{(1)}_{(\pm)}-iJ^{(2)}_{(\pm)}\right) \bar\epsilon^{\pm}D_\pm\bbX{}\right),
\eer{add}
Identifying this with the reduced form of \re{UVdefOne} using \re{proj} we find
\footnote{As stated in \cite{Goteman:2012qk} this identification is  up to a phase corresponding to $R$ symmetry of the $\theta$s. A different choice of phase can interchange the identification of $J^{(1)}$ and $J^{(2)}$.}
\ber
&&\half\left(J^{(1)}_{(\alpha)}-iJ^{(2)}_{(\alpha)}\right)=U^{(\alpha)}\pi_{(\alpha)}~,\cr
&&~~~\cr
&&\half\left(J^{(1)}_{(\alpha)}+iJ^{(2)}_{(\alpha)}\right)=V^{(\alpha)}\bar\pi_{(\alpha)}~.
\eer{cmp}
%\rd{Both sides nilpotent}\\
It can be  verified that this agrees with \re{Jmult} and the other properties of of $U^{(\pm)}$ and $V^{(\pm)}$ and it leads to the expected \((1,1)\) identification as follows:\\
From \re{cmp} we have that
\ber\nn
&&J^{(1)}_{(\alpha)}=U^{(\alpha)}\pi_{(\alpha)}+V^{(\alpha)}\bar\pi_{(\alpha)}\\[1mm]
&&J^{(2)}_{(\alpha)}=i\left(U^{(\alpha)}\pi_{(\alpha)}-V^{(\alpha)}\bar\pi_{(\alpha)}\right).
\eer{49}
Since \(\pi\) and  \(\bar\pi\) are orthogonal projection operators we find, using \re{pUp} and \re{VU}
\ber\nn
(J^{(1)}_{(\alpha)})^2&\!=\!&U^{(\alpha)}\pi_{(\alpha)}U^{(\alpha)}\pi_{(\alpha)}+U^{(\alpha)}\pi_{(\al)}V^{(\al)}\bar\pi_{(\al)}\\[1mm]\nn
&&
+V^{(\al)}\bar\pi_{(\alpha)}U^{(\alpha)}\pi_{(\alpha)}+V^{(\alpha)}\bar\pi_{(\alpha)}V^{(\alpha)}\bar\pi_{(\alpha)}\\[1mm]
&=& U^{(\alpha)}V^{(\alpha)}\bar\pi_{(\alpha)}+V^{(\alpha)}U^{(\alpha)}\pi_{(\alpha)}
\eer{}
It is further easy to verify that together with $(J^{(3)}_{(\alpha)})$ the complex structures $(J^{({\cal A})}_{(\alpha)})$ satisfy \re{SU2}.
Now \(UV\) and \(UV\) are projection operators \re{info}. They preserve the  
\(\pi\)s.  We illustrate this for $U^{(+)}V^{(+)}\bar\pi^{(+)}$ and $U^{(+)}V^{(+)}\pi^{(+)}$. We have
\ber
U^{(+)}V^{(+)}=-\left(\begin{array}{cccc}
\one &0&0&0\\ 
0&0&0&0\\
0&0&\one&0\\
0&0&0&\one
\end{array}\right)~,~~~V^{(+)}U^{(+)}=-\left(\begin{array}{cccc}
0 &0&0&0\\ 
0&\one&0&0\\
0&0&\one&0\\
0&0&0&\one
\end{array}\right)
\eer{}
From \re{Jpm} and \re{projo} we see that
\ber
\bar\pi=
\left(\begin{array}{cccc}
\one &0&0&0\\ 
0&0&0&0\\
\star&\star&\star&\star\\
\star&\star&\star&\star
\end{array}\right)~,~~~
\pi=
\left(\begin{array}{cccc}
0&0&0&0\\ 
0&\one&0&0\\
\star&\star&\star&\star\\
\star&\star&\star&\star
\end{array}\right)
\eer{}
so that $UV\bar\pi=-\bar\pi$ and $VU\pi=-\pi$ and
\ber
(J^{(1)}_{(+)})^2=-\bar\pi-\pi=-\one
\eer{}

The condition \re{inv} that follows from invariance of the $(2,2)$ action \re{Asemis1} will correspond to the usual preservation of the metric (see \re{covj});
\ber
(J_{(+)}^{({\cal A})})^tg=gJ_{(+)}^{({\cal A})}
\eer{}
and covariant constancy of the complex 
structures 
\ber
\na^{(\pm)}_{\pm}J^{({\cal A})}_{(\pm)}=0~,
\eer{}
that follow from the invariance of the $(2,2)$ action \re{Asemis1}, reduced to $(1,1)$
\ber
S=\int D^2D_+\bbX{i}_|E_{ij}D_-\bbX{j}_|~,
\eer{}
under the transformations 
\ber
\delta^{{\cal A}} \bbX{i}_|=\epsilon^{+}_{{\cal A}} J^{i({\cal A})}_{(+)j}D_+ \bbX{j}_|+\epsilon^{-}_{{\cal A}} J^{i({\cal A})}_{(-)j}D_- \bbX{j}_|~,
\eer{}
where
\ber\nn
&&\epsilon^{\al}_{1}=\epsilon^{\al} + \bar \epsilon^{\al}\\
&&\epsilon^{\al}_{2}=i(\epsilon^{\al} - \bar \epsilon^{\al})
\eer{}
in \re{add}.
It is clear from the identification \re{cmp} and \re{49} that these conditions hold for $U^{(\al)}\pi$ and $V^{(\al)}\bar \pi$, so for the transformations \re{UVdefOne}.

Note that the projection operators \re{projo}, $\pi_{(\al)}$, for each $\al$, split the  (complexified) tangent bundle into two maximally isotropic sub-bundles according to
\ber
T{\cal M}\oplus \mathbb{C}={\mathbb{L}}^{(1,0)}_{(\al)}\oplus{\mathbb{L}}^{(0,1)}_{(\al)}
\eer{}
where ${\mathbb{L}}^{(1,0)}_{(\al)}$ is the $+i$ and ${\mathbb{L}}^{(0,1)}_{(\al)}$ the $-i$ eigenbundle of $J_{(\al)}$. The transformations \re{YFtfs} now read
\ber
\left(\begin {array}{c}
\bar\delta^{(\alpha)}X\cr
\delta^{(\alpha)}Y\end{array}\right)
=\left(\begin {array}{cc}
0&\half\left(J^{(1)}_{(\alpha)}- iJ^{(2)}_{(\alpha)}\right)\cr
\half\left(J^{(1)}_{(\alpha)}+iJ^{(2)}_{(\alpha)}\right)&0\end{array}\right)\left(\begin {array}{c}
\epsilon^{(\alpha)}D_\al X\cr
\bar\epsilon^{(\alpha)}D_\al Y
\end{array}\right)~,
\eer{}
where $X\in{\mathbb{L}}^{(1,0)}_{(\al)}$ and $Y\in {\mathbb{L}}^{(0.1)}_{(\al)}$ and the $F$ structures squares to $-\one$ on $({\mathbb{L}}^{(1,0)}_{(\al)},{\mathbb{L}}^{(0,1)}_{(\al)})^t$.\\

This concludes the \((1,1)\)  identification of the additional supersymmetry for the  semichirals.

\section{Discussion}
In this paper it has been shown how nonmanifest supersymmetries realised on the semichiral fields of a symplectic sigma model lead to a new type of geometry consisting of two Yano $F$-structures and {an almost} para hermitian geometry on the doubled tangent space. This represents a novel way of constructing supersymmetric sigma models where one or more extra symmetries do not correspond directly to complex structures. The $F$-structures do become  complex structures on  sub-bundles, so that the geometry may alternatively be described in terms of CR structures, i.e., there is an involutive sub-bundle which carries a complex structure.

In defining integrability for the $F$-structure on the doubled tangent space the relation between double field theory and Generalised Geometry was mimicked closely . The whole set-up may presumably be recast into structures on the generalised tangent bundle. One indication of this is that we may perform a coordinate transformation on  the coordinates of one tangent bundle, $\bbY{i}$ say,
\ber
\bbY{i}\to \bbY{}_i:=K,_i
\eer{}
which would give coordinates much like the momentum and winding mode coordinates. Showing that this preserves the $F$-structures the usual transition to Generalised Geometry \cite{Hitchin:2003cxu} could follow. So it would lead to a Generalised Geometry description of the  $F$-structures.\\

 {\bf Acknowledgments}\\
  I thank A. Avranitakis, P.S. Howe,  %C.M. Hull  
  and M. Ro\v cek for discussions and comments.
  The research  is 
supported in part by the 2236 Co-Funded Scheme2 (CoCirculation2) of T\"UB{\.I}TAK 
(Project No:120C067)\footnote{\tiny However 
the entire responsibility for the publication is ours. The financial support received from 
T\"UB{\.I}TAK does not mean that the content of the publication is approved in a scientific 
sense by T\"UB{\.I}TAK.}. \\
\eject
\appendix

{\bf \Large Appendix}

\section{Index conventions}
The two dimensional spinor indicex $\al$ takes the values $(+)$ and $(-)$, and the vector indices are the pairs $\+$ and $=$. When in parenthesis, $(\al)$ is a label, not an index.
\bigskip

\noindent{\bf $(2,2)$ conventions:}\\

The $(2,2)$ algebra of covariant spinor derivatives is 
\ber
\{\bbD{\pm},\bbDB{\pm} \}=2i\partial_{\pp}~,
\eer{}
all other anticommuting.
Left $L$, and right $R$ semichiral superfields are denoted 
\ber
L =(\ell^a,\bar\ell^{\bar a})^t~, ~\makebox{and}~  R =(\scriptr^{a'},\bar{\scriptr}^{\bar a'})^t
\eer{}
and obey
\ber
\bbDB{+}\ell^a=0~,~ \makebox{and} ~\bbDB{-}\scriptr^{a'}=0~,
\eer{}
where $a,\bar a, a', \bar a'=1\dots d$~.
Collectively they are denoted
\ber
\bbX{i}= (L,R)~,
\eer{}
where $i=1\dots 4d$. They are coordinates on ${\cal M}$. When doubling the space we double these coordinates $X^i\to (\bbX{i},\bbY{j})$ so that 
\ber
\bbX{I}=(\bbX{i},\bbY{j})~,
\eer{}
with $I=1\dots 8d$, are coordinates on $\mathbb{M}$. In \re{domg}, which preceedes the doubling discussion, $(z^i, \bar z^i)$ replace $(\bbX{i},\bbY{j})$.\\

\noindent
$(\bbX{}, \bbD{+}\bbX{},  \bbDB{+}\bbX{})$ are coordinates on the doubled tangent bundle.

\eject

\noindent{\bf $(1,1)$ conventions:}\\

The $(2,2)$ algebra is 
\ber
D_{\pm}^2=Q_{\pm}^2=i\partial_{\pp}~,
\eer{}
all other anticommuting. The reduced superfields are
$\bbX{i}_|$ and the $\theta$-projection is then dropped.
The bi-quaternionic set of complex structures is 
\ber
(J^{({\cal A})}_{(\al)})~,
\eer{}
where ${\cal A} = 1,2,3$.

\section{Hermiticity}
A summary of some properties of semichirals.\\

\noindent
For the symplectic  model the sum of the metric and $B$-field read \cite{Lindstrom:2005zr}
\ber\nn
&&E_{LL} = C_{LL}K^{LR}J_s K_{RL}{}\\[1mm] \nonumber
&&E_{LR} = J_s K_{LR} J_s + C_{LL}K^{LR}C_{RR }{}\\[1mm] \nonumber
&&E_{RL} = -K_{RL} J_s K^{LR}J_s K_{RL}{}\\[1mm]
&&E_{RR} = -K_{RL} J_s K^{LR}C_{RR} ~.
\eer{Efield}
The symmetric part of $E$ is
\ber
&&G_{LL}=C_{LL}K^{LR}JK_{RL}-K_{LR}JK^{RL}C_{LL}\cr
&&~~\cr
&&G_{LR}=JK_{LR}J+C_{LL}K^{LR}C_{RR}-K_{LR}JK^{RL}JK_{LR}\cr
&&~~\cr
&&G_{RR}=-K_{RL}JK^{LR}C_{RR}+C_{RR}K^{RL}JK_{LR}~,
\eer{metric}
and the antisymmetric part is
\ber
&&B_{LL}=C_{LL}K^{LR}JK_{RL}+K_{LR}JK^{RL}C_{LL}\cr
&&~~\cr
&&B_{LR}=JK_{LR}J+ C_{LL}K^{LR}C_{RR }+K_{LR}JK^{RL}JK_{LR}\cr
&&~~\cr
&&B_{RR}=K_{RL}JK^{LR}C_{RR}+C_{RR}K^{RL}JK_{LR}~.
\eer{Bfield}
This agrees with \re{MetB}. 
\ber\nonumber
&&G=\Omega[J_+,J_-]{}\\[1mm]
&&B=\Omega\{J_+,J_-\}
\eer{altdef2}
using \re{Jpm}.
It follows that
\ber\nn
G_{Li}J^i_{(+)L}&=&C_{LL}K^{LR}JK_{RL}J+JK_{LR}JK^{RL}C_{LL}
+C_{LL}K^{LR}C_{RR}K^{RL}C_{LL}\\[1mm]\nn
G_{Li}J^i_{(+)R}&=&-G_{Ri}J^i_{(+)L}=JK_{LR}JK^{RL}JK_{LR}+C_{LL}K^{LR}C_{RR}K^{RL}JK_{LR}+K_{RL}J\\[1mm]
G_{Ri}J^i_{(+)R}&=&-C_{RR}-K_{RL}JK^{LR}C_{RR}K_{RL}JK_{LR}
\eer{}
From this, hermiticity wrt $J_{(+)}$ is manifest
\ber
\omega_{(+)ij}=G_{ik}J^k_{(+)j}=-\omega_{(+)ji}
\eer{}
The corresponding  calculation for $J_{(-)}$ shows that the model is bi-hermitian.


\begin{thebibliography}{99}

%\cite{Zumino:1979et}
\bibitem{Zumino:1979et}
B.~Zumino,
``Supersymmetry and Kahler Manifolds,''
Phys. Lett. B \textbf{87} (1979), 203
doi:10.1016/0370-2693(79)90964-X
%688 citations counted in INSPIRE as of 18 Jul 2022

%\cite{Alvarez-Gaume:1980xat}
\bibitem{Alvarez-Gaume:1980xat}
L.~Alvarez-Gaume and D.~Z.~Freedman,
``Ricci Flat Kahler Manifolds and Supersymmetry,''
Phys. Lett. B \textbf{94} (1980), 171-173
doi:10.1016/0370-2693(80)90850-3
%93 citations counted in INSPIRE as of 18 Jul 2022

%\cite{Gates:1984nk}
\bibitem{Gates:1984nk}
S.~J.~Gates, Jr., C.~M.~Hull and M.~Ro\v cek,
``Twisted Multiplets and New Supersymmetric Nonlinear Sigma Models,''
Nucl. Phys. B \textbf{248} (1984), 157-186
doi:10.1016/0550-3213(84)90592-3

%\cite{Hull:1985jv}
\bibitem{Hull:1985jv}
C.~M.~Hull and E.~Witten,
``Supersymmetric Sigma Models and the Heterotic String,''
Phys. Lett. B \textbf{160} (1985), 398-402
doi:10.1016/0370-2693(85)90008-5
%383 citations counted in INSPIRE as of 18 Jul 2022]

%\cite{Hull:2017hfa}
\bibitem{Hull:2017hfa}
C.~Hull and U.~Lindstr\"om,
``All $(4,0)$: Sigma Models with $(4,0)$ Off-Shell Supersymmetry,''
JHEP \textbf{08} (2017), 129
doi:10.1007/JHEP08(2017)129
[arXiv:1707.01918 [hep-th]].
%4 citations counted in INSPIRE as of 18 Jul 2022

%\cite{Buscher:1987uw}
\bibitem{Buscher:1987uw}
T.~Buscher, U.~Lindstr\"om and M.~Ro\v cek,
``New Supersymmetric $\sigma$ Models With {Wess-Zumino} Terms,''
Phys. Lett. B \textbf{202} (1988), 94-98
doi:10.1016/0370-2693(88)90859-3

%\cite{Lindstrom:1994mw}
\bibitem{Lindstrom:1994mw}
U.~Lindstr\"om, I.~T.~Ivanov and M.~Ro\v cek,
``New N=4 superfields and sigma models,''
Phys. Lett. B \textbf{328} (1994), 49-54
doi:10.1016/0370-2693(94)90426-X
[arXiv:hep-th/9401091 [hep-th]].
%36 citations counted in INSPIRE as of 18 Jul 2022


%\cite{Hull:1985pq}
\bibitem{Hull:1985pq}
C.~M.~Hull, A.~Karlhede, U.~Lindstr\"om and M.~Ro\v cek,
``Nonlinear $\sigma$ Models and Their Gauging in and Out of Superspace,''
Nucl. Phys. B \textbf{266} (1986), 1-44
doi:10.1016/0550-3213(86)90175-6




%\cite{Goteman:2009xb}
\bibitem{Goteman:2009xb}
M.~G\"oteman and U.~Lindstr\"om,
``Pseudo-hyperkahler Geometry and Generalized Kahler Geometry,''
Lett. Math. Phys. \textbf{95} (2011), 211-222
doi:10.1007/s11005-010-0456-7
[arXiv:0903.2376 [hep-th]].

%\cite{Lindstrom:2014bra}
\bibitem{Lindstrom:2014bra}
U.~Lindstr\"om,
``Extended supersymmetry of semichiral sigma models in 4D,''
JHEP \textbf{02} (2015), 170
doi:10.1007/JHEP02(2015)170
[arXiv:1411.3906 [hep-th]].

%\cite{Goteman:2012qk}
\bibitem{Goteman:2012qk}
M.~G\"oteman, U.~Lindstr\"om and M.~Ro\v cek,
{\em Semichiral Sigma Models with 4D Hyperkaehler Geometry,}
JHEP \textbf{01} (2013), 073
doi:10.1007/JHEP01(2013)073
[arXiv:1207.4753 [hep-th]].

\bibitem{lost}
U. Lindstr\"om, I. Ryb, M. Ro\v cek, R.von Unge, M. Zabzine, {\em T-duality for the $S^1$ piece in the  $S^3xS^1$ model}, October 2009, unpublished.

%\cite{Goteman:2009ye}
\bibitem{Goteman:2009ye}
M.~G\"oteman, U.~Lindstr\"om, M.~Ro\v cek and I.~Ryb,
``Sigma models with off-shell N=(4,4) supersymmetry and noncommuting complex structures,''
JHEP \textbf{09} (2010), 055
doi:10.1007/JHEP09(2010)055
[arXiv:0912.4724 [hep-th]].



\bibitem{Yano:1961}
K.~Yano,
{\it On a structure f satisfying $f^3+f=0$}, Tech. Rep. Univ. of Washington, {\bf 12} (1961); 
{\it On a structure defined by a tensor field of type $(1,1)$ satisfying $f^3+f=0$},
Tensor, N.S. {\bf 14}, 9 (1963).

\bibitem{IshiharaYano}
S. Ishihara and K. Yano,
{\it On integrability conditions of a structure $f$ satisfying $f^3+f=0$},
Quart. J. Math. {\bf 15} (1964) 217-222.
		
%\cite{Lindstrom:2005zr}
\bibitem{Lindstrom:2005zr}
U.~Lindstr\"om, M.~Ro\v{c}ek, R.~von Unge and M.~Zabzine,
{\em Generalized K\"ahler manifolds and off-shell supersymmetry},
Commun.\ Math.\ Phys.\  {\bf 269} (2007) 833
[arXiv:hep-th/0512164].
%%CITATION = CMPHA,269,833;%%

%\cite{Kimura:2022jyp}
\bibitem{Kimura:2022jyp}
T.~Kimura, S.~Sasaki and K.~Shiozawa,
``Complex structures, T-duality and worldsheet instantons in Born sigma models,''
JHEP \textbf{06} (2022), 119
doi:10.1007/JHEP06(2022)119
[arXiv:2203.03272 [hep-th]].
%1 citations counted in INSPIRE as of 07 Jul 2022

\bibitem{Whitney}
Weisstein, Eric W. "Whitney Sum." From MathWorld--A Wolfram Web Resource. https://mathworld.wolfram.com/WhitneySum.html

%\cite{Hitchin:2003cxu}
\bibitem{Hitchin:2003cxu}
N.~Hitchin,
``Generalized Calabi-Yau manifolds,''
Quart. J. Math. \textbf{54} (2003), 281-308
doi:10.1093/qjmath/54.3.281
[arXiv:math/0209099 [math.DG]].
%625 citations counted in INSPIRE as of 22 Jul 2022



\end{thebibliography}
\end{document}